# A Dynamic Tree Algorithm for On-demand Peer-to-peer Ride-sharing Matching

**Rui Yao**

Department of Civil and Environmental Engineering,

Technion – Israel Institute of Technology, Haifa 32000, Israel

Email: andyyao@campus.technion.ac.il

**Shlomo Bekhor**

Department of Civil and Environmental Engineering,

Technion – Israel Institute of Technology, Haifa 32000, Israel

Email: sbekhor@technion.ac.il

## ABSTRACT

Innovative shared mobility services provide on-demand flexible mobility options and have the potential to alleviate traffic congestion. These attractive services are challenging from different perspectives. One major challenge in such systems is to find suitable ride-sharing matchings between drivers and passengers with respect to the system objective and constraints, and to provide optimal pickup and drop-off sequence to the drivers.

In this paper, we develop an efficient dynamic tree algorithm to find the optimal pickup and drop-off sequence. The algorithm finds an initial solution to the problem, keeps track of previously explored feasible solutions, and reduces the solution search space when considering new requests. In addition, an efficient pre-processing procedure to select candidate passenger requests is proposed, which further improves the algorithm performance.

Numerical experiments are conducted on a real size network to illustrate the efficiency of our algorithm. Sensitivity analysis suggests that small vehicle capacities and loose excess travel time constraints do not guarantee overall savings in vehicle kilometer traveled.

Yao, Bekhor

# INTRODUCTION

Traffic congestion is one of the major issues that many cities are facing today. Both public agencies and private companies are trying to develop efficient mobility options. Innovative shared mobility services provide on-demand flexible mobility options and have the potential to alleviate traffic congestion. One interesting form is peer-to-peer ridesharing, in which ride-sharing drivers need to perform activities other than only serving passengers. Table 1, adapted from Masoud and Jayakrishnan (2017), compares different service alternatives.

**Table 1 Comparison of different service alternatives**
(adapted from Masoud and Jayakrishnan, 2017)

| Service components | | Taxi | Taxi-sharing | Ride-sourcing | Ride-splitting | peer-to-peer ride-sharing | para-transit (disabled, dial-a-ride) | Informal Carpooling |
|---|---|---|---|---|---|---|---|---|
| Vehicle | Company Fleet | ✓ | ✓ | | | | ✓ | |
| | Self-own vehicle | | | ✓ | ✓ | ✓ | | ✓ |
| Drivers | Employee | ✓ | ✓ | | | | ✓ | |
| | Freelance driver | | | ✓ | ✓ | | | |
| | Peer driver | | | | | ✓ | | ✓ |
| Arrangement | Pre-arranged | ✓ | ✓ | ✓ | ✓ | ✓ | ✓ | ✓ |
| | On-demand | ✓ | ✓ | ✓ | ✓ | ✓ | ✓ | |
| Multiple Passengers | | | ✓ | | ✓ | ✓ | ✓ | ✓ |
| Cost sharing | | | ✓ | | ✓ | ✓ | ✓ | ✓ |

The emergence of peer-to-peer ride-sharing services brings a variety of challenges. One major challenge is the ride-sharing matching, which optimally assigns passengers to drivers with respect to the system objective and constraints, and provides optimal pickup and drop-off sequence to the drivers.

The problem to determine the optimal pickup and drop-off sequence for ride-sharing drivers is well-known as the dial-a-ride problem (DARP). Extensive literature in DARP focused on algorithms for pre-arranged dial-a-ride services. However, in a realistic ride-sharing setting, ride-sharing requests arrive continuously and are not known in advance. Most existing DARP solution algorithms search for a new solution from scratch when new requests appear (Cordeau, 2006; Braekers et al., 2014; Gschwind and Drexl, 2019), ignoring the previously explored feasible solutions.

Recent works have developed efficient many-to-many algorithms in the context of real-time applications. Masoud and Jayakrishnan (2017) introduced the ellipsoid spatiotemporal accessibility method (ESTAM) to reduce the search space and solved the matching problem using dynamic programing. Alonso-Mora *et al.* (2017) proposed a general many-to-many dynamic multi-passenger vehicle assignment framework by reducing the problem to a passenger-combination (or trip) to driver matching problem by shareability networks and solved using integer linear program (ILP). In Simonetto et al. (2019), the ride-sharing matching problem is simplified to a single-passenger assignment problem and solved using linear programming.

This paper presents a mathematical formulation of on-demand ride-sharing matching problem and develops an efficient dynamic tree algorithm to solve problem. The proposed dynamic tree algorithm is a many-to-many algorithms, in which multiple passengers are considered and assigned to drivers at the same time. The contributions of the present paper are outlined as follows:

1. We provide a mathematical formulation for the ride-sharing matching problem, which also considers service quality constraints (e.g. waiting time and maximum riding time).
2. We propose a geometric pruning procedure to efficiently eliminate the infeasible passenger requests for each driver based on spatiotemporal proximity of both drivers and passengers.
3. We adapt a dynamic tree structure to solve the ride-sharing DARP. The algorithm keeps track of the feasible solutions for a given set of passenger requests using a dynamic tree. New requests are inserted into these feasible solutions with respect to constraints. Our algorithm is a generalization of Huang et al., (2017) in the sense that any number of passengers can be assigned to a vehicle (provided there are available seats).
4. We perform analysis on algorithm performance and impact of ride-sharing on the network.

**PROBLEM FORMULATION**

Different objectives were studied in the literature. Here, we formulate the ride-sharing matching problem as an optimization problem with a societal objective to provide a convenient ride-sharing service and reduce vehicle kilometer travel (VKT). The formulation is based on a passenger-driver network, where each node represents either origin $o$ or destination $d$ of drivers and passengers, and arc $(a,b) \in A$ represents the shortest path between node $a$ and $b$, resulting in an objective function:

$$\max z = \left(\sum_{r_j \in R} l(o_j, d_j) + \sum_{v_i \in V} l(o_i, d_i)\right) - \sum_{v_i \in V} \sum_{(a,b) \in A} l_{(a,b)} \cdot x_{(a,b)}^{v_i} \qquad (1)$$

where $\sum_{r_j \in R} l(o_j, d_j)$ and $\sum_{v_i \in V} l(o_i, d_i)$ are the total shortest path distances for passengers $R$ and drivers $V$, respectively. And $\sum_{v_i \in V} \sum_{(a,b) \in A} l_{(a,b)} \cdot x_{(a,b)}^{v_i}$ is the total kilometer traveled with ridesharing, with $x_{(a,b)}^{v_i}$ being the decision variable if passenger is matched with driver. The constraints of this optimization problem are discussed as follows. Constraint (2) indicates that the number of picked up passengers for a driver is equal to drop-off passengers for a request $r_j$ at his/her origin and destination.

$$q_{o_j} + q_{d_j} = 0 , \forall r_j \in R \qquad (2)$$

Constraint (3) is the flow conservation: it ensures that a driver arriving at a node in the network will also leave the node at the same time period. Moreover, it ensures that if the passenger is picked up, the driver will also have to drop him off at his destination.

$$\sum_{a:(a,b)\in A} x^{v_i}_{(a,b)} = \sum_{c:(b,c)\in A} x^{v_i}_{(b,c)} = z^{v_i}_{r_b}, \forall b \in N_R, \forall v_i \in V \qquad (3)$$

Eq. (4) shows that if route of driver $v_i$ passes through arc $(a,b)$, then the arrival time at node $b$ is equal to the sum of arc travel time $tt_{(a,b)}$ and the arrival time at node $a$.

$$t_{v_i}(b) = \sum_{a:(a,b)\in A} x^{v_i}_{(a,b)} \cdot \left(tt_{(a,b)} + t_{v_i}(a)\right), \forall b \in N \qquad (4)$$

Constraint (5) defines how the occupancy $Q_{v_i}$ changes at each node the driver visits:

$$Q_{v_i}(b) = \sum_{a:(a,b)\in A} x^{v_i}_{(a,b)} \cdot \left(q_b + Q_{v_i}(a)\right), \forall b \in N \qquad (5)$$

Eq. (6) indicates that each node in the passenger-driver network is visited at most once, i.e. a pickup or drop-off is performed at most once:

$$\sum_{v_i \in V} \sum_{a:(a,b)\in A} x^{v_i}_{(a,b)} \leq 1, \forall b \in N_R \qquad (6)$$

Eq. (7) ensures that every driver will leave his/her origin and arrive at his/her destination:

$$\sum_{b:(o_i,b)\in A} x^{v_i}_{(o_i,b)} = \sum_{a:(a,d_i)\in A} x^{v_i}_{(a,d_i)} = 1, \forall v_i \in V \qquad (7)$$

Constraint (8) guarantees that each destination node will only be visited after its paired origin is visited:

$$t_{v_i}(d_k) \leq t_{v_i}(o_k), \forall v_i \in V, \text{where } k \in \{v_i, r_j | r_j \in R, z^{v_i}_{r_j} = 1\} \qquad (8)$$

Eq. (9) assures that if passengers are assigned to a driver, each pickup node can only be visited after the earliest departure time:

$$t_{v_i}(o_k) \geq t_{ED}(k), \forall v_i \in V, \text{where } k \in \{v_i, r_j | r_j \in R, z^{v_i}_{r_j} = 1\} \qquad (9)$$

For simplicity, we also assume drivers depart from their origins at their earliest departure times, $t_{v_i}(o_i) = t_{ED}(v_i)$. Constraints (10)-(11) indicate level-of-service constraints. The maximum waiting time $\Omega_{r_j}$ is set in eq. (10), and the maximum excess travel time $\Delta_k$ is captured by eq. (11), where $\tau(o,d)$ is the shortest path travel time.

$$\sum_{v_i \in V} \left(t_{v_i}(o_j) - t_{ED}(r_j)\right) \cdot z^{v_i}_{r_j} \leq \Omega_{r_j}, \forall r_j \in R \qquad (10)$$

$$\sum_{v_i \in V} \left(t_{v_i}(d_j) - t_{ED}(r_j) - \tau(o_j, d_j)\right) \cdot z^{v_i}_{r_j} \leq \Delta_k, \forall k \in (V \cup R) \qquad (11)$$

The last constraint, eq. (12) ensures at any node on driver's route, vehicle capacity $c_i$ is never violated, where all vehicles are initialized with no passenger, $Q_{v_i}(o_i) = 0$:

$$Q_{v_i}(a) \leq c_i, \forall v_i \in V, \text{where } a \in \{o_i, d_i, o_j, d_j | r_j \in R, z^{v_i}_{r_j} = 1\} \qquad (12)$$

Yao, Bekhor## SOLUTION APPROACH

We follow the common practice in literature to decouple the matching problem, by extending the state-of-the-art framework proposed by Alonso-Mora et al. (2017) with a simple pre-processing procedure and an efficient dynamic tree algorithm to obtain service sequences, and result a bi-level solution framework (Figure 1).

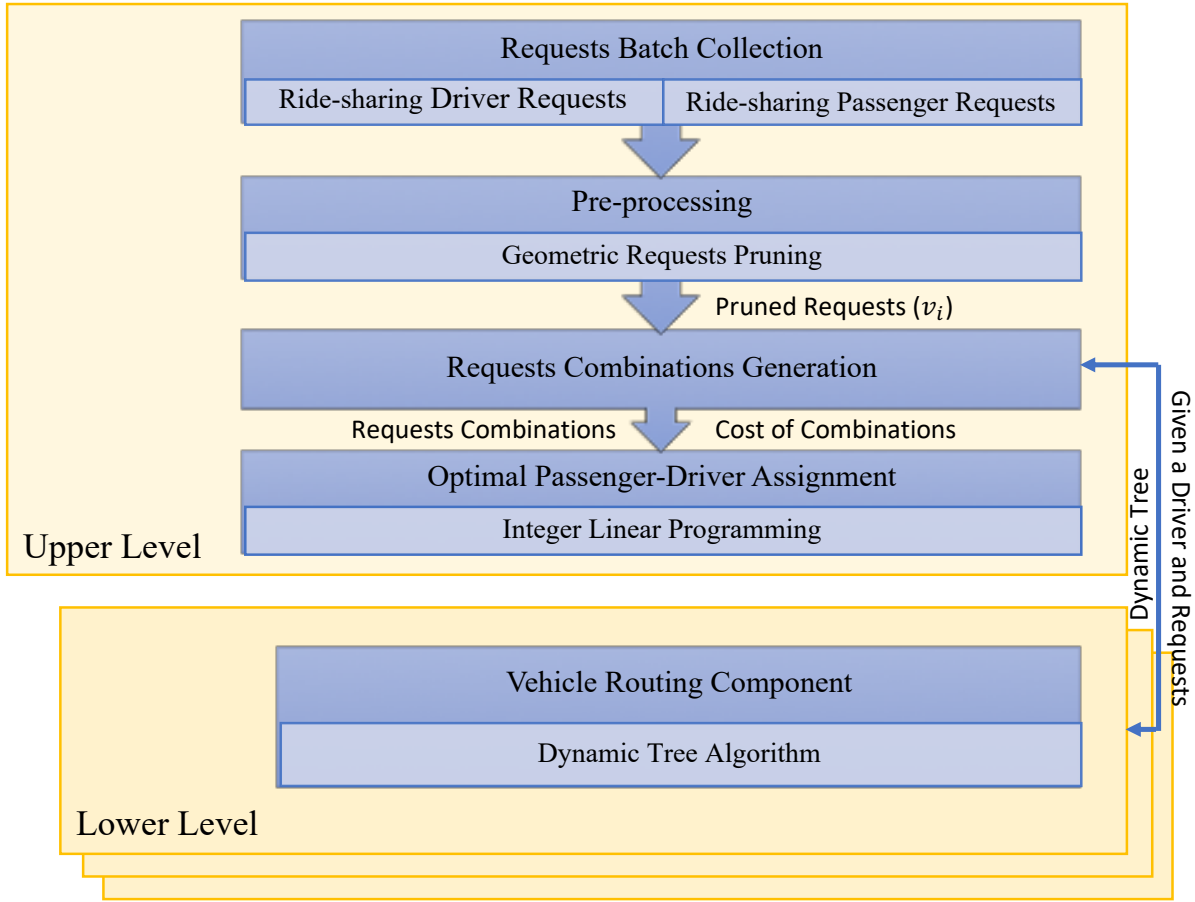

Figure 1: Solution framework for the On-demand ride-sharing problem

The ride-sharing matching solution framework contains mainly four steps. First, the passenger and driver requests are grouped respectively for a predefined period of time. Then the Pre-processing step efficiently extracts the candidate passenger requests. Next, the Request Combinations Generation step applies Dynamic Tree Algorithm to combine the feasible passenger requests and to find the driver's service sequences. The last step corresponds to the solution of the optimal passenger-driver assignment problem using integer linear program (ILP). The following describes in detail the main algorithm steps.

Yao, Bekhor

**Pre-processing – Geometric Requests Pruning**

The main idea of the pre-processing step is to use the pickup time window constraints and maximum driving time constraints to select the candidate passengers-driver pairs that may satisfy these constraints.

We defined the accessible region of the driver as an ellipse, where the focal points are his origin and destination, and transvers diameter is related to the maximum driving time. The reachable pickup region of the passenger is defined as a circle, where the center is the pickup location and radius relates to the latest pickup time. This step simply finds passengers and drivers that intersect each other as the candidate passenger set. Figure 2 shows an example of the pruning process with 2 drivers and 2 passenger requests, and the final candidate passenger is shown in (c).

The candidate passenger-driver pairs are passed to the Request Combinations Generation step, to form passenger combinations that are feasible to be served by the driver. The feasibility of the passenger combinations is checked by finding a service sequence that satisfies both driver and passengers constraints. Since many combinations are needed to be checked, an efficient algorithm is required, which is developed in the next section.

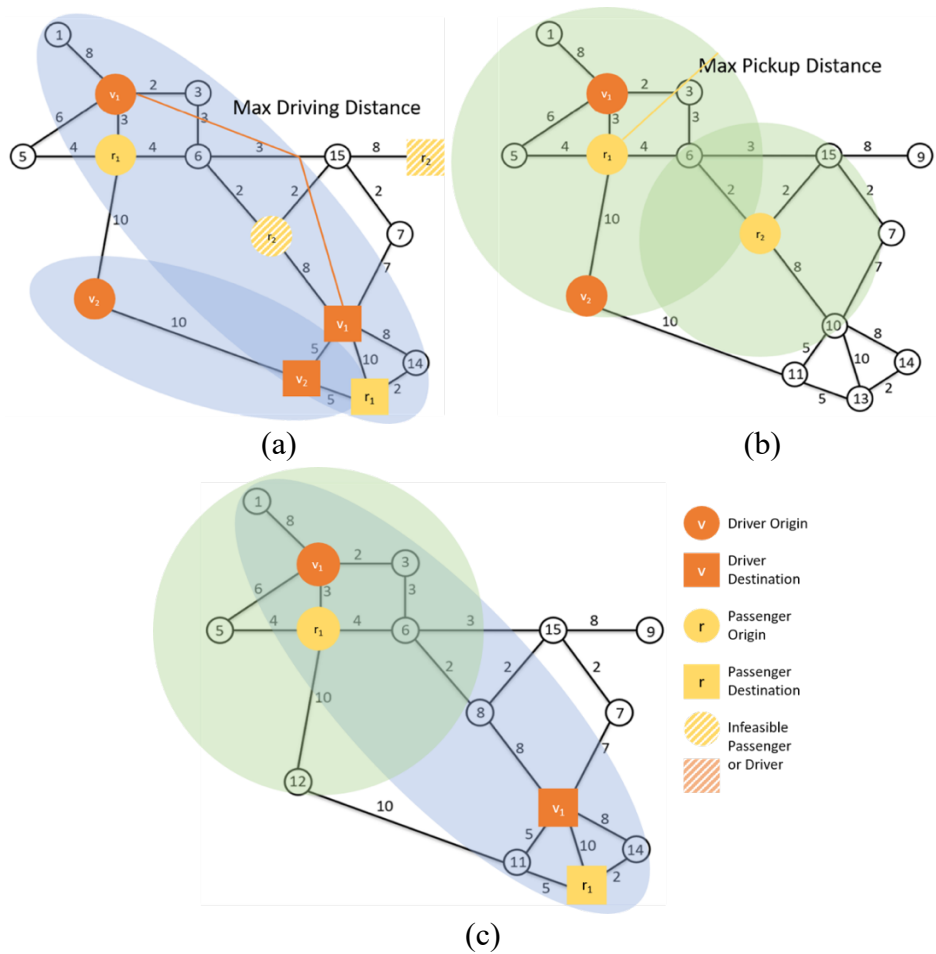

**Figure. 2, Pruning Example**

**Dynamic Tree Algorithm**

We propose a dynamic tree algorithm for finding optimal ride-sharing pickup and drop-off service sequence. The fundamental idea of the algorithm is to keep track of the feasible solutions for a given set of passenger requests, using a dynamic tree structure. Because infeasible solutions of the previous set of passenger requests will remain infeasible when considering service sequence with the new requests, we only need to insert the new requests into the existing dynamic tree to obtain new feasible solutions.

Speedup heuristics derived from solution properties of DARP are introduced to further reduce the solution search space. If any time constraint is violated at any location when inserting into the dynamic tree, the algorithm can stop the procedure because the time constraints will remain violated for the consecutive nodes in the dynamic tree. However, if capacity constraint is violated, it depends on whether the insertion is for pickup or drop-off. If it is a pickup node, the procedure can search for insertion after some drop offs occur. On the other hand, the capacity constraint needs to be valid for any node between a pair of pickup and drop-off nodes.

Figure 3 outlines the proposed dynamic tree algorithm. The algorithm requires input of a driver's current feasible service sequences in terms of a dynamic tree T, the insertion node list N=(o,d) where o and d is the pickup and drop-off node respectively.

The algorithm contains three main procedures: REQUEST INSERTION, TREE EXPANSION and CASE DETERMINATION. We will first insert the origin and construct a new tree that is feasible to incorporate the pickup. The next step will be inserting the drop-off node in the dynamic tree after the pickup to ensure the precedence constraint. The algorithm will start from the root of T and moves downwards in T, to explore all feasible solutions with the new request. The following example will explain the algorithm steps in more detail.

Yao, Bekhor

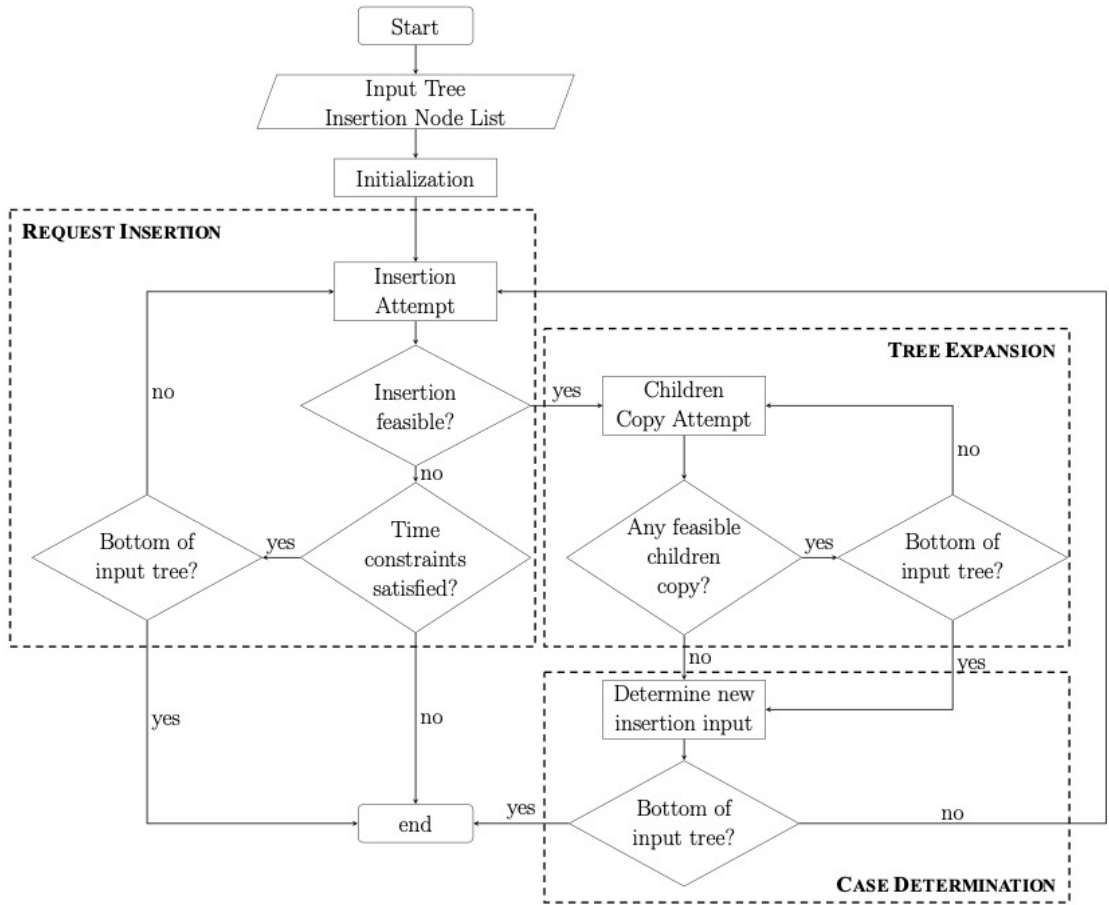

**Figure. 3, Dynamic Tree Algorithm Flowchart**

*Example*

Consider an example that a driver $v_1$ is willing to take 2 passengers, and there is already a passenger $r_1$ onboard, the algorithm will try to find new service sequences to include the new request $r_2$. The corresponding initial input tree $T$ is shown in Figure 4(a), notice that there are 3 possible insertion points for origin node for $r_2$.

First, REQUEST INSERTION procedure tries to insert the origin node for $r_2$ at insertion point 1. For the given vehicle capacity and number of requests, the capacity constraint is never violated, thus for simplicity, capacity constraint check is ignored. Suppose the earliest pickup time and maximum waiting time constraint of origin node for $r_2$ at insertion point 1 are satisfied, we first copy the origin node for $r_2$ to a new tree $T'$ as shown in Figure 4(b).

At the end of each procedure, CASE DETERMINATION decides the next procedure based on the inserted node type and feasibility conditions. In this example, CASE DETERMINATION decides to check if the constraints of existing nodes are still satisfied, using the TREE EXPANSION procedure.

The TREE EXPANSION procedure recursively copies nodes (dashed line box in Figure 4(a)) from initial input tree $T$ to the new tree under origin node for $r_2$ in $T'$, and verify all the constraints.

Assume that the additional waiting time for pickup endured by $r_1$ is still within his/her maximum waiting time constraint, it is feasible to copy the origin node for $r_1$ to $T'$. However, due to the extra detour to pick up $r_2$, the maximum excess travel time constraint for $r_1$ is violated, so it is infeasible to copy the destination node for $r_1$ to $T'$. The TREE EXPANSION procedure is terminated due to violation of time constraints (indicated by the crossing lines in Figure 4(c)), meaning that, pickup $r_2$ before picking $r_1$ is not feasible, and we will move down to insert origin node for $r_2$ at point 2.

Now, suppose the REQUEST INSERTION of origin node for $r_2$ at insertion point 2 is feasible and there is a feasible new tree after the TREE EXPANSION procedure. The resulting new tree $T'$ is shown in Figure 4(d), this means we found a feasible service sequence that is able to incorporate picking up $r_2$, and satisfying all constraints for both the driver and request $r_1$. The next step will be finding feasible insertions for the destination node of $r_2$.

The destination node for $r_2$ is inserted under the origin node for $r_2$ in the new tree $T'$, in which there are 2 potential insertion points (shown in Figure 4(d)). The destination node for $r_2$ is inserted firstly at point 4. Assume the maximum excess travel time constraint is satisfied for destination node for $r_2$ at point 4. The algorithm attempts to copy the remaining nodes in the tree, in this case only the driver's destination $d_v$, into $T'$. The same process is repeated for insertion point 5, and assume all the constraints are satisfied, a feasible new tree $T'$ with respect to insertion of origin node for $r_2$ at point 2 is obtained as shown in Figure 4(e). At this point, we already found 2 feasible pickup and drop-off sequences that are able to handle the new request $r_2$.

Lastly, suppose the insertion of origin node for $r_2$ is also feasible at insertion point 3, and all the constraints are satisfied for the destination node for $r_2$ and driver's destination $d_v$ under pickup node $o_2$ at insertion point 3. The dynamic tree algorithm found 3 feasible service sequences to accommodate the new request $r_2$ (Figure 4(f)). The optimal sequence and its objective value will be given to the upper level request combination step.

Yao, Bekhor

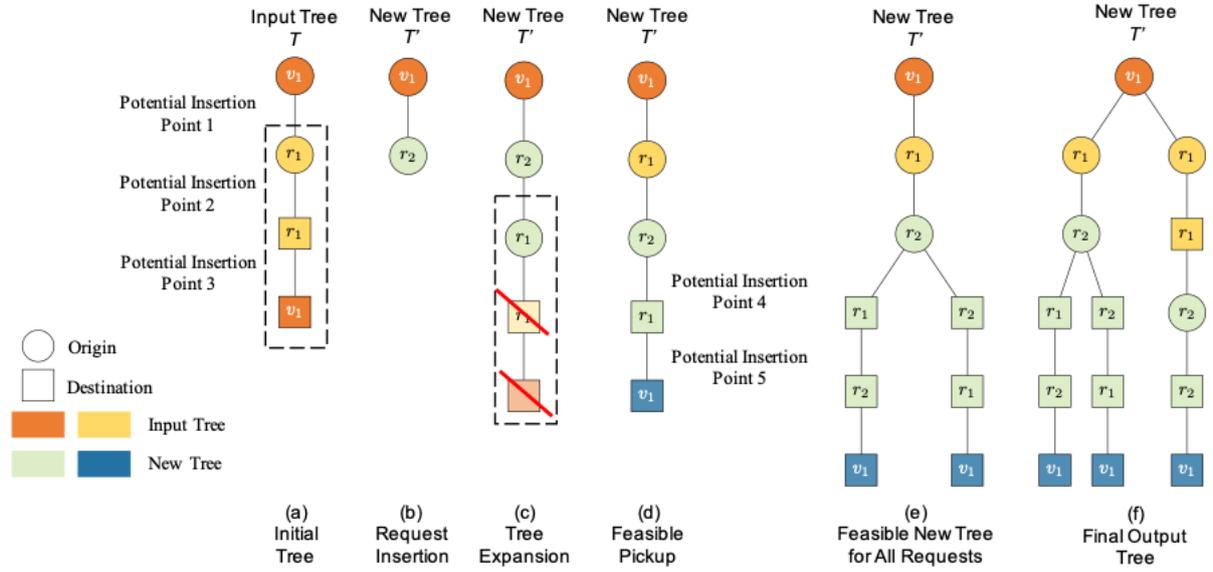

**Figure 4, Example of Dynamic Tree Algorithm steps**

**Request Combinations Generation**

After running the dynamic tree algorithm, the feasibility of the passenger combination – driver pair can be verified. With this feasibility information, this step creates passenger combinations based on cliques in graph theory.

It first checks all the requests provided by the Pre-Processing, check if the driver can serve each one of these requests alone. Next, for all the requests that can be served by the driver alone, we check if the combination of the 2 requests can be served by the driver with respect to the constraints. Continue checking for combining 3, 4, ..., k requests by joining the feasible combinations with size (k-1).

An greedy variant of the Dynamic Tree is applied in this step to speed up the calculations. Instead of using the whole Dynamic Tree, we only insert the new request to the optimal branch with the best objective value.

The complexity of the Dynamic Tree Algorithm is $O\left(\frac{(2m)!}{2^m} \cdot (2m+1)\right) = O((2m)!)$ in the worst case, and the heuristic version of the Dynamic Tree Algorithm is of order $O(m^2 \cdot (2m+1)) = O(m^3)$.

Yao, Bekhor

**Optimal passenger-driver assignment**

As the last step of the solution framework, the original complex integer program is simplified to be a bipartite matching problem:

$$\max z = C \cdot X \quad (13)$$

$$\sum_{j=1}^{n} \varphi(i,j) \cdot x(j) \leq 1, \forall i \in [1, |R|] \quad (14)$$

$$\sum_{j=1}^{n} \psi(i,j) \cdot x(j) \leq 1, \forall i \in [1, |V|] \quad (15)$$

$$x(j) \in \{0,1\}, \forall j \in [1, n] \quad (16)$$

where $C$ is a vector of objective value of passenger combination – driver pair obtained from request combination step, and $x \in X$ is a vector of decision variables which equals to 1 if the request combination is matched with the driver, and 0 otherwise. Constraints 14 and 15 ensure each passenger and driver is matched at most once, where $\varphi$ and $\psi$ represent the passenger combination – driver pair incidents.

**RESULTS**

We exemplify the algorithm performance using the well-known Winnipeg network. The network consists of 154 zones, 1,067 nodes, 2,535 links, 4,345 origin-destination (OD) pairs, and a total demand of 54,459 trips. We assume 3,000 ride-sharing participant requests, in which 1,000 of them are willing to be drivers, and the remaining 2,000 are passenger requests. Ride-sharing vehicles have capacity of 4 and are empty at the beginning of the time period. The maximum excess travel time constraints for passengers and drivers are assumed as 20% of their shortest path travel time. The maximum pickup waiting time is assumed as 50% of the maximum excess travel time.

100 replications with random sample of OD trip distribution were performed (Figure 5), in which the algorithm runtime is almost linear to the total number of feasible request combinations. The average total runtime with pruning is 49.02 sec, and average runtime without pruning process is 63.64 sec, the pruning process save 15 second runtime at a cost of 0.1 second.

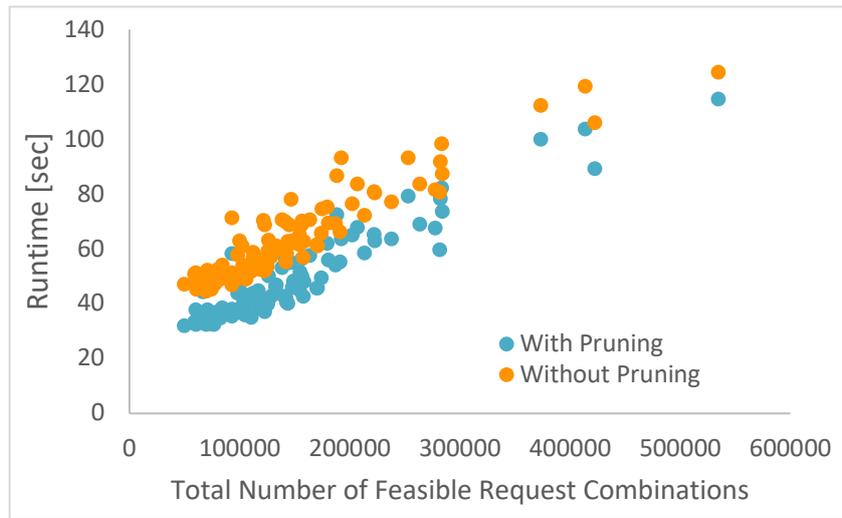

**Figure 5 Total runtimes for each of the 100 random draws**

The impact of ride-sharing on the network is also evaluated, results are summarized in Table 2. Note that, the VKT and Vehicle-hour travelled (VHT) savings are relatively smaller than vehicle trips savings. This is because we allow drivers detour to pick up and drop off passengers.

Table 2, Summary of Ride-sharing Impact on the Network

|  | Base Case | Average (100 Replications) | Difference |
|---|---|---|---|
| Vehicle Trips | 54,459 | 53,291 | 2.14% |
| VKT | 720,902 | 706,047 | 2.06% |
| VHT | 706,567 | 692,223 | 2.03% |

We conducted sensitivity analysis on different supply-demand relations (under supply, balance, and over supply), vehicle capacities and excess travel time constraints. One of the important results is that, negative VKT saving is observed in ride-sharing systems under supply and with loose excess travel time constraints (Figure 6). This suggests that the ride-sharing system should be carefully designed to maintain its sustainability and network benefits.

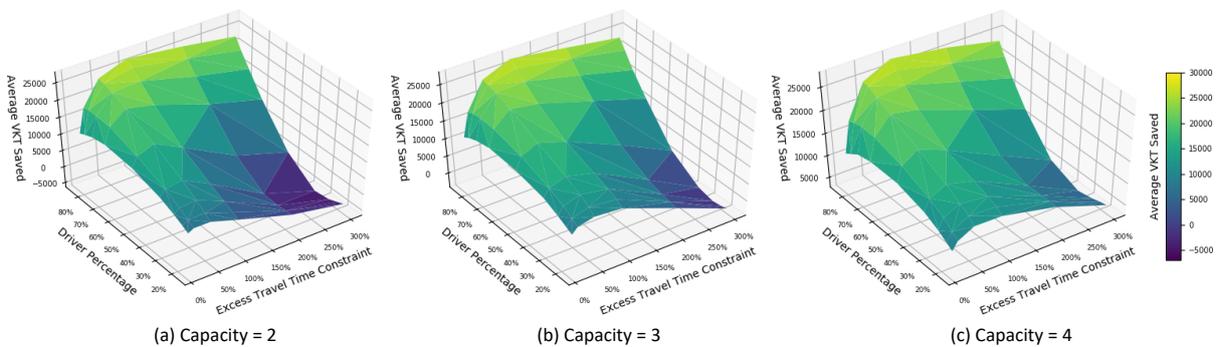

(a) Capacity = 2    (b) Capacity = 3    (c) Capacity = 4

**Figure 6 Sensitivity analysis on VKT savings**

**SUMMARY AND OUTLOOK**

In this paper, we propose an efficient dynamic tree algorithm with simple and effective pre-processing step to solve the on-demand ride-sharing matching problem. One of our unique contributions is to integrate the dynamic tree algorithm for solving ride-sharing VRP. The dynamic tree provides the algorithm the ability to utilize the previously calculated driver schedules, instead of calculating them from scratch.

The numerical tests confirm that our proposed dynamic tree algorithm is computational efficient. Another contribution of this paper is the inclusion of a simple geometric pruning procedure to reduce the size of the problem. Consequently, runtime results for the Pre-processing step are very small, and this procedure helps reduce the overall algorithm runtimes.

This paper exemplifies the algorithm implementation using the well-known Winnipeg network. We performed initial tests related to OD distributions, driver supply, excess travel time constraints and vehicle capacities. Numerical experiments show that the spatial distribution of ride-sharing participants influences the algorithm performance. Sensitivity analysis confirm that the most critical ride-sharing matching constraints are the excess travel times.

The network analysis using traffic assignment suggests that ride-sharing system should be carefully designed to maintain the sustainability of a ride-sharing system, and help alleviate the traffic congestion in the network. In particular, the results suggest that small vehicle capacities do not guarantee overall VKT savings.

This paper assumed several parameters related to the demand-side constraints, such as maximum excess travel times, maximum waiting times, and earliest pickup times. The matching results are of course dependent on these parameters. Further research is needed to provide recommendations with respect to ride-sharing level of service.

The excess travel times were calculated by using a fixed travel time matrix. The re-routing of drivers may change the travel times on the network (because of the change in traffic flows), and consequently may change the matching results. Further research will investigate the stability of the matching results with respect to travel times.